\documentclass[fleqn,10pt,onecolumn]{SelfArx} 
\usepackage{graphicx}
\PassOptionsToPackage{hyphens}{url}\usepackage{hyperref}
\usepackage{subfigure}
\usepackage{listings}
\usepackage[square,numbers]{natbib}
\definecolor{color1}{RGB}{0,0,90}  
\definecolor{color2}{RGB}{0,20,20} 

\usepackage{hyperref} 
\hypersetup{hidelinks,colorlinks,breaklinks=true,urlcolor=color2,citecolor=color1,linkcolor=color1,bookmarksopen=false,pdftitle={Title},pdfauthor={Author}}

\JournalInfo{\mbox{\;}} 
\Archive{} 

\PaperTitle{Benchmarking the computing resources at the Instituto 
            de Astrof{\'\i}sica de Canarias} 

\Authors{
    Nicola Caon\textsuperscript{1,2}*, 
    Antonio J. Dorta\textsuperscript{1,2}, 
    Juan Carlos Trelles Arjona\textsuperscript{2}
}

\affiliation{\textsuperscript{1}\textit{Instituto de Astrof{\'\i}sica de Canarias, 
              E-38205 La Laguna, Tenerife, Spain}} 
\affiliation{\textsuperscript{2}\textit{Universidad de La Laguna, 
              Dpto. Astrof{\'\i}sica, E-38206 La Laguna, Tenerife, Spain}} 
\affiliation{*\textbf{Corresponding author}: nicola.caon@iac.es} 

\newcommand{\keywordname}{Keywords} 

\Abstract{ 
The aim of this study is the characterization of the computing resources used
by researchers at the "Instituto de Astrof{\'\i}sica de Canarias" (IAC).
Since there is a huge demand of computing time and we use tools such as
HTCondor to implement High Throughput Computing (HTC) across all available PCs, 
it is essential for us to assess in a quantitative way, using
objective parameters, the performances of our computing nodes.
In order to achieve that, we have run a set of benchmark tests on a number of
different desktop and laptop PC models among those used in our institution. 
In particular, we run the "Polyhedron
Fortran Benchmarks" suite, using three different compilers: GNU Fortran 
Compiler, Intel Fortran Compiler and the PGI Fortran Compiler; 
execution times are then
normalized to the reference values published by Polyhedron. The same tests
were run multiple times on a same PCs, and on 3 to 5 PCs of the same model
(whenever possible) to check for repeatability and consistency of the
results. We found that in general execution times, for a given PC model, 
are consistent within an uncertainty of about 10\%, and show a gain in 
CPU speed of a factor of about 3 between the oldest PCs used at the IAC 
(7-8 years old) and the newest ones.
}

\begin{document}

\def\UrlFont{\sl}
\onecolumn

\flushbottom 

\maketitle 

\tableofcontents 

\thispagestyle{empty} 


\section{The IAC computing resources} 


At the Instituto de Astrof{\'\i}sica de Canarias there are about 250 desktop
PCs with Linux installed, used by scientists and engineers. These PCs cover a
wide span of models and ages, from 8-years old Dell Optiplex to recently
bought "NausicaA" models.
There are also several more powerful computers, mainly rack-mounted but also a
few desktop models, dedicated to large, demanding jobs that exceed the
capabilities of a "regular" (consumer) PC, such as massive data reduction and 
analysis, simulations, and other CPU intensive jobs.

While it is clear that newer PCs are faster and more efficient than older
models, so far this was more a perception than solid evidence supported by 
data. A user may observe that her office-mate's latest-model PC is more 
responsive, or faster when executing some tasks, but cannot say by how much,
nor can she estimates the gain in time obtained by running her  
applications in the office-mate's PC instead of her own older PC.
This could be a key factor when preparing the remote executions 
of a program using the available HTC tools, since it is possible to specify
a list of preferences or ranks that will be used to choose the target machines
on which the code will be executed.

For these reasons we decided to run a set of benchmarking tests on all the
different, available desktop and rack models, with also a few
laptops, as part of a month-long "Proyecto Pr{\'a}ctica de Empresa" (Student
Internship), carried out at the IAC by a 4th-year student of Astronomy
(JCTA)\footnote{The tests were run in November/December 2014, with some
additional runs in January-March 2015}.

\section{Running the benchmark tools} 

\subsection{The Polyhedron benchmarks suite}

After considering a number of possible benchmarks, we finally selected  the
"Polyhedron Fortran Benchmarks" suite 
\cite{MFL6VW74649}, since it is one of the most
comprehensive set of benchmarks that matches our requirements: it provides
tools to automatically run the tests, compute the CPU time used by each
executable, validate and save the results in tabular form in a simple
text file.

This benchmark consists of 17 independent Fortran programs. While this suite
was devised to compare the performances of 10 different Fortran compilers on a
same machine, it can be used as well to compare the performances of a same
compiler on a variety of hardware.

The way it operates is controlled by a couple of parameter files, one
(general) listing the tests to be run, the desired accuracy to be achieved,
the minimum and maximum number of runs for each test, and the maximum 
execution time permitted. (A detailed explanation of how the test suite
works is provided in \cite{benchinstall}).
Then there is a parameter file tailored to each compiler, with the specific
command and flags to be used.

We made only minor changes in the parameter files provided by Polyhedron, by
increasing the tolerance on the execution time from 0.1 to 0.2,
setting the maximum number of runs for each test to 20, and limiting the
maximum execution time to 4000 seconds.
These changes do not affect the reliability of our results, but allow a
shorter overall time for running the whole suite of tests (typically from two
to seven hours). The limits on the maximum number of tests and maximum 
execution time (per test) prevented jobs from getting out of control and
using up the CPU for hours or even days (which happened a few times,
especially with the Intel compiler).

To check the consistency of the results, we: \newline
a) ran the benchmarks multiple times on a same machine, and  \newline
b) ran the tests on 3 to 5 different PCs of a same desktop model. 
However, we typically have only one single model of the more powerful machines
dedicated to CPU-intensive jobs, so we could only perform the consistency check a).
This same limitation applies to laptops.

Table~\ref{Table:listofPCs} lists the hardware on which the Polyhedron test
suite was run, together with the main data about their CPU and RAM.

Table~\ref{Table:compilers} lists the compilers installed at the IAC and
used for the benchmark tests.
All the computers on which the tests were run had Linux Fedora 19 
installed, and all have the same exact version of the three compilers used.

\begin{table*}
\begin{footnotesize}
\begin{tabular}{llllrrc}
\hline 
\textbf{PC model}       &  \textbf{Type}     &  \textbf{Date}       & \textbf{Processor Type}      & \textbf{Cache} & RAM   & \textbf{Number} \\  
\hline\\[-6pt]
Dell Precision WS T7400          &  Desktop  & late 2007   & Intel(R) Xeon(R) CPU X5472  @ 3.00GHz          &   6144 KB     &  32 GB & 2  \\  
Dell Optiplex 740$^\mathrm{(a)}$ &  Desktop  & late 2007   & AMD Athlon(tm) 64 X2 Dual Core Processor 5200+ &   1024 KB     &   4 GB & 5  \\  
Dell Optiplex 740$^\mathrm{(a)}$ &  Desktop  & early 2008  & AMD Athlon(tm) 64 X2 Dual Core Processor 5600+ &    512 KB     &   4 GB & 4  \\  
Dell Optiplex 740$^\mathrm{(a)}$ &  Desktop  & mid 2009    & AMD Athlon(tm) 64 X2 Dual Core Processor 6000+ &   1024 KB     &   4 GB & 5  \\  
Dell Optiplex 780$^\mathrm{(b)}$ &  Desktop  & late 2009   & Intel(R) Core(TM)2 Quad CPU Q9400 @ 2.66GHz    &   3072 KB     &   8 GB & 8  \\  
Dell Optiplex 780$^\mathrm{(b)}$ &  Desktop  & late 2009   & Intel(R) Core(TM)2 Duo CPU E8400 @ 3.00GHz     &   3072 KB     &   8 GB & 4  \\  
Dell Precision WS T3500          &  Desktop  & early 2011  & Intel(R) Xeon(R) CPU W3565  @ 3.20GHz          &   8192 KB     &  12 GB & 1  \\  
Dell Precision WS T3600          &  Desktop  & mid 2012    & Intel(R) Xeon(R) CPU E5-1650 0 @ 3.20GHz       &  12288 KB     &  32 GB & 1  \\  
Dell Precision WS T5600          &  Desktop  & mid 2012    & Intel(R) Xeon(R) CPU E5-2687W 0 @ 3.10GHz      &  20480 KB     & 128 GB & 1  \\  
Dell Optiplex 7010               &  Desktop  & mid 2012    & Intel(R) Core(TM) i5-3470 CPU @ 3.20GHz        &   6144 KB     &   8 GB & 5  \\  
ALDA+                            &  Desktop  & mid 2014    & Intel(R) Core(TM) i7-4770 CPU @ 3.40GHz        &   8192 KB     &   8 GB & 5  \\  
NausicaA                         &  Desktop  & mid 2014    & Intel(R) Core(TM) i7-4790 CPU @ 3.60GHz        &   8192 KB     &   8 GB & 4  \\  
Dell Precision WS T5400          &  Rack     & late 2007   & Intel(R) Xeon(R) CPU X5450  @ 3.00GHz          &   6144 KB     &   8 GB & 1  \\  
Dell Precision WS-690            &  Rack     & early 2008  & Intel(R) Xeon(R) CPU X5355  @ 2.66GHz          &   4096 KB     &  32 GB & 2  \\  
Dell PowerEdge-2970              &  Rack     & late 2008   & Quad-Core AMD Opteron(tm) Processor 2358 SE    &    512 KB     &  64 GB & 1  \\  
Dell PowerEdge R410              &  Rack     & late 2009   & Intel(R) Xeon(R) CPU X5570  @ 2.93GHz          &   8192 KB     &  16 GB & 1  \\  
Tecal RH5885 V3                  &  Rack     & mid 2014    & Intel(R) Xeon(R) CPU E7-4820 v2 @ 2.00GHz      &  16384 KB     & 256 GB & 1  \\  
Dell Latitude E6500              &  Laptop   & mid 2008    & Intel(R) Core(TM)2 Duo CPU T9400  @ 2.53GHz    &   6144 KB     &   4 GB & 1  \\  %
Dell Latitude E4200              &  Laptop   & late 2008   & Intel(R) Core(TM)2 Duo CPU U9600 @ 1.60GHz     &   3072 KB     &   3 GB & 1  \\  
Dell Latitude E4300              &  Laptop   & late 2008   & Intel(R) Core(TM)2 Duo CPU P9300  @ 2.26GHz    &   6144 KB     &   4 GB & 1  \\  
Dell Latitude E6320              &  Laptop   & early 2011  & Intel(R) Core(TM) i7-2720QM CPU @ 2.20GHz      &   6144 KB     &   4 GB & 1  \\  
Dell Latitude E6520              &  Laptop   & early 2011  & Intel(R) Core(TM) i7-2620M CPU @ 2.70GHz       &   4096 KB     &   8 GB & 2  \\  
Lenovo L440                      &  Laptop   & late 2014   & Intel(R) Core(TM) i5-4200M CPU @ 2.50GHz       &   3072 KB     &   8 GB & 1  \\[1pt] 
%
%
%
\hline\\[2pt] 
\end{tabular}
\caption{Column "Date" show the production date of that specific model as retrieved from 
the corresponding Dell page \cite{Dell} after supplying the service
tag. For some of the models the Dell website does not provide useful information, so
we listed the approximate time the PC model was bought (this also applies to
non-Dell models). \newline
Within each "Type" group, the list is ordered chronologically. \\
$^\mathrm{(a)}$: The Dell Optiplex 740 model actually came with three CPU variants;  
$^\mathrm{(b)}$: The Dell Optiplex 780 model actually came with two CPU variants.   
Column "Number" shows the number of PCs of that model on which the benchmark tests were
run.}
\label{Table:listofPCs}
\end{footnotesize}
\end{table*}

\begin{table*}
\begin{small}
\begin{tabular}{lll}
\hline
\textbf{compiler}  &  \textbf{version}                                     &  \textbf{compilation flags}                     \\  
\hline
gfortran       &  GNU Fortran (GCC) 4.8.3 20140911                         &  -march=native -ffast-math -funroll-loops -O3   \\
Intel          &  ifort (IFORT) 14.0.2 20140120                            &  -O3 -fast -parallel -ipo -no-prec-div          \\
PGI            &  pgf90 14.10-0 64-bit target on x86-64 Linux -tp penryn   &  -V -fastsse -Munroll=n:4 -Mipa=fast,inline     \\[1pt]
\hline\\[-6pt]
\end{tabular}
\caption{Compiler version and flags used for the benchmarks tests. Our
compiler versions are slightly different from those used in the Polyhedron
Suite (listed as gfortran 4.9, Intel 15.0, PGI 14.9).}
\label{Table:compilers}
\end{small}
\end{table*}

\subsection{Using HTCondor to manage the benchmark jobs}

Ideally, the benchmarks tests should be run on a dedicated machine, with no
other processes running, in order to minimize the CPU load and guarantee that
the results reflect the best performances the hardware can deliver.

However, we could not afford to take PCs away from their users, so the tests
were run on production PCs, i.e. PC used (generally during the day) by their
users. So we had a twofold problem: on one side, we wanted our tests not to
interfere with the usage of the PCs by their users; on the other side, we did
not want to run the tests on a PC with a high CPU load which can obviously 
affect the results. 
 
HTCondor provides a nice and efficient solution to this problem. HTCondor is a
distributed job scheduler developed by the University of Wisconsin-Madison,
which allows users to run their applications in other users' machines when
they are not being used (for details about HTCondor, see
\cite{htcondor_url, htcondor_paper}).

We first made a initial selection of machines where to run the tests,
choosing whenever possible, among all the available desktop models, 
those we knew were less heavily used. This information was gathered
by using \textit{ConGUSTo} \cite{congusto}, a tool that provides 
real-time and historical usage data about the machines forming the 
HTCondor pool.
Based on these data, the final list comprises about 60 PCs.

In order not to run two or more benchmark instances on a same PC (HTCondor
tries to use all the available "slots", that is CPU cores), we restricted our
jobs to run only on "slot1". The list of target machines was included in the
requirements of our HTCondor submit files.

HTCondor only runs its jobs on those PCs that are not being used and
that have a CPU load below a certain threshold (so with no CPU- or 
memory-heavy background jobs). If the CPU load rises, or the user goes back
to work interactively, the HTCondor job is killed and rescheduled for 
the next available opportunity (on any of the target machines).

Thus the first thing we did was to submit via HTCondor a batch of
benchmarks jobs (each job is the complete suite of tests for a specific
compiler) to all the targets machines.

If the required number of benchmarks executions was obtained for a specific
PC, it was removed from the machines target list and a batch of HTCondors 
jobs was submitted again. A few iterations were generally sufficient to
complete the benchmark runs on most PCs, while for a few of them it was
necessary to prepare and submit HTCondor jobs restricting the targets list to
that specific PC.

In all PCs used for the benchmarks Hyper-threading was disabled. Moreover, as
all the benchmarks run sequentially on just one single core of the CPU, the 
results do not depend on how many core are in the CPU.

Listing ~\ref{lst:htcondor} is an example of the HTCondor submission files we
used, with detailed comments about the various settings and commands.


\lstset{
        frame=single,
        numbersep=5pt,
        commentstyle=\itshape,
        basicstyle=\scriptsize,
        keywordstyle=\bfseries,
        sensitive=false,
        morecomment=[l]{\#},
        keywords={output,error,log,universe,should\_transfer\_files,when\_to\_transfer\_output,getenv,transfer\_input\_files,transfer\_output\_files,transfer\_output\_remaps,requirements,executable,arguments,queue}
}
\begin{lstlisting}[caption={HTCondor submit file},label={lst:htcondor}]
#####################################################
# HTCondor Submit File
#####################################################

# 1) Number of executions, filenames. 
# The benchmark test is run N times.
# FNAME is the basename of output files.
N     = 5 
ID    = $(Cluster).$(Process)
FNAME = condor_pb11

# 2) Managing stdout, stderr and logs
output  = $(FNAME).$(ID).out
error   = $(FNAME).$(ID).err    
log     = $(FNAME).$(Cluster).log

# 3) Basic settings. Enable file transfer, so that
# output files are copied from the PC where the job
# is run to the PC which manage submissions and 
# store all output files in a central location.
# getenv will pass the environment on to the HTCondor 
# job.
universe                = vanilla
should_transfer_files   = YES 
when_to_transfer_output = ON_EXIT
getenv                  = True

# 4) Transfer files. Only the runtimes.txt file, the
# one storing the execution times of the various tests, 
# is relevant (bldtimes lists the times required to 
# compile the tests, while exesizes lists the size of 
# the executables).
BASE                   = .../pb11/lin
transfer_input_files   = $(BASE)
transfer_output_files  =               \
          $(BASE)/source/bldtimes.txt, \ 
          $(BASE)/source/exesizes.txt, \ 
          $(BASE)/source/runtimes.txt, \ 
          $(BASE)/source/pgi_118_lin_SB.sum

# 5) Rename output files so as to include the name of
# the PC in which the test was run for easy 
# identification.
transfer_output_remaps  = \ 
    "runtimes.txt=runtimes--$$(NAME)--$(ID).txt;..."

# 6) Force executions only in slot1, so that no two or
# more benchmark tests are run simultaneously on a same
# PC. String "machines" list the name of the PCs on 
# which we want to run the HTCondor jobs.
slot         = substr(toLower(Target.Name), 0, 6)
machines     = "mach1,mach2,mach3,...,machN"
requirements = ($(slot) == "slot1@"))  &&           \
         stringListMember(UtsnameNodename,$(machines))

# 7) Executable and arguments (in our case all
# relevant flags and arguments are included in the
# bash script)
executable = $(BASE)/Condor/bnchmrk-Polyhedron-pgi.bash
arguments  = ""

# 8) Submit the job to the queue!
queue $(N)

\end{lstlisting}



\subsection{Consistency checks}

\begin{figure*}
\includegraphics[angle=0,width=\linewidth]{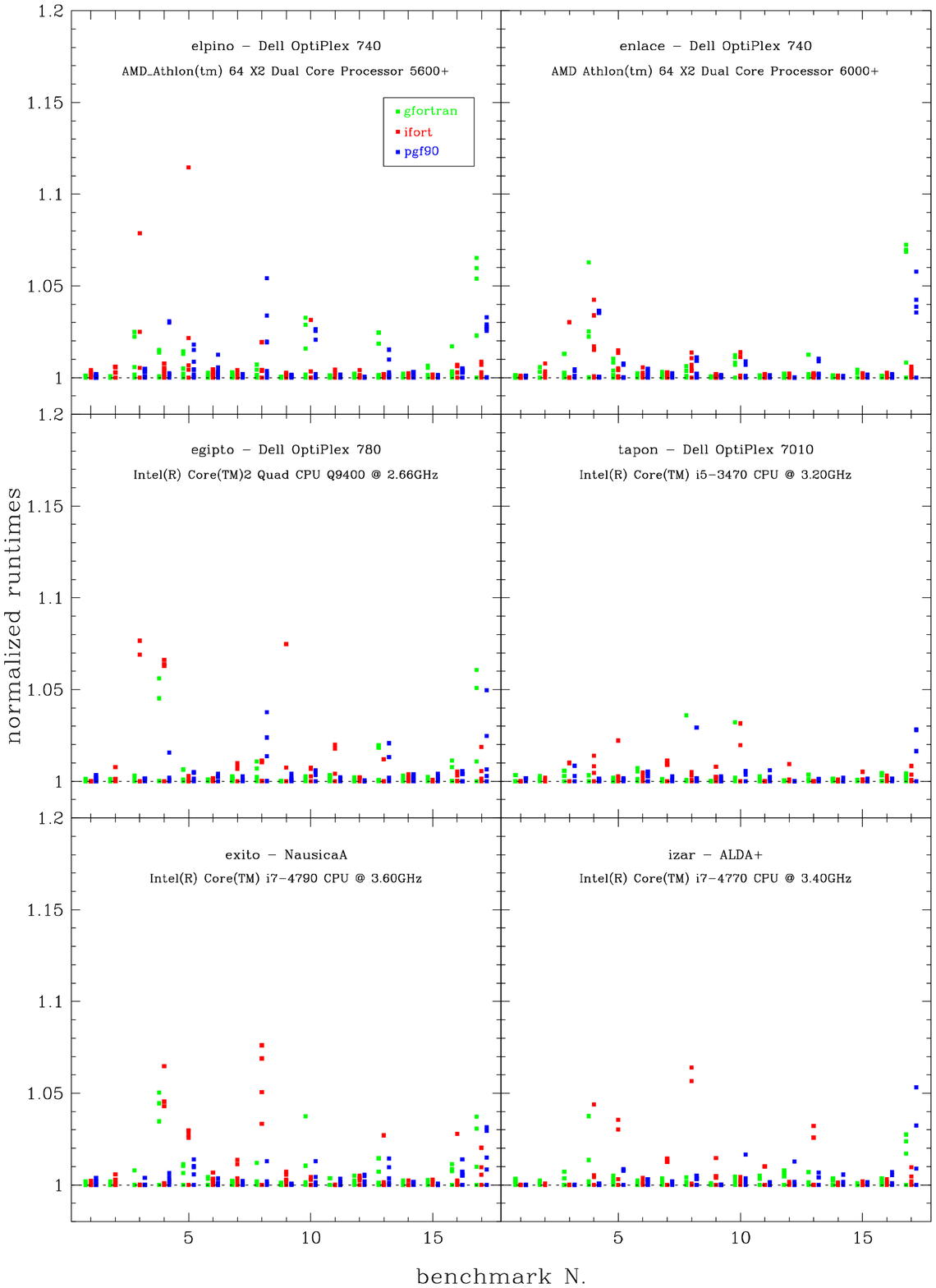}
\caption{This panel shows, for six different individual PCs, the scatter in
run-times after normalizing them to the best (minimum) value for each
benchmark test and compiler. Green points refer to the \textsl{gfortran}
compiler, red points to \textsl{ifort},  blue points to \textsl{pgf90}. The
benchmark test N. 2 (\textsl{aermod}) fails when compiled with
\textsl{pgf90}, and is thus omitted.}
\label{Fig:comparison_same_machine}
\end{figure*}

A minimum of 3, and up to 9 runs per compiler and per PCs were obtained
in order to check for consistency and repeatability of the results.
We found that, for a same machine, the execution times varied within a 
few percentage points. We then took the minimum value for each test as
our final result for each PC and compiler.
Figure~\ref{Fig:comparison_same_machine} illustrates a few example of 
how the execution times vary across the various runs in a same PC.

In a few cases some runs produced weird results, with execution times much
higher than expected (often only for just some specific tests). For
some reason, this happened more frequently with the Intel compiler. Those
runs were excluded, and new runs submitted if necessary to meet the minimum
number of runs we set.

\begin{figure*}
\includegraphics[angle=0,width=\linewidth]{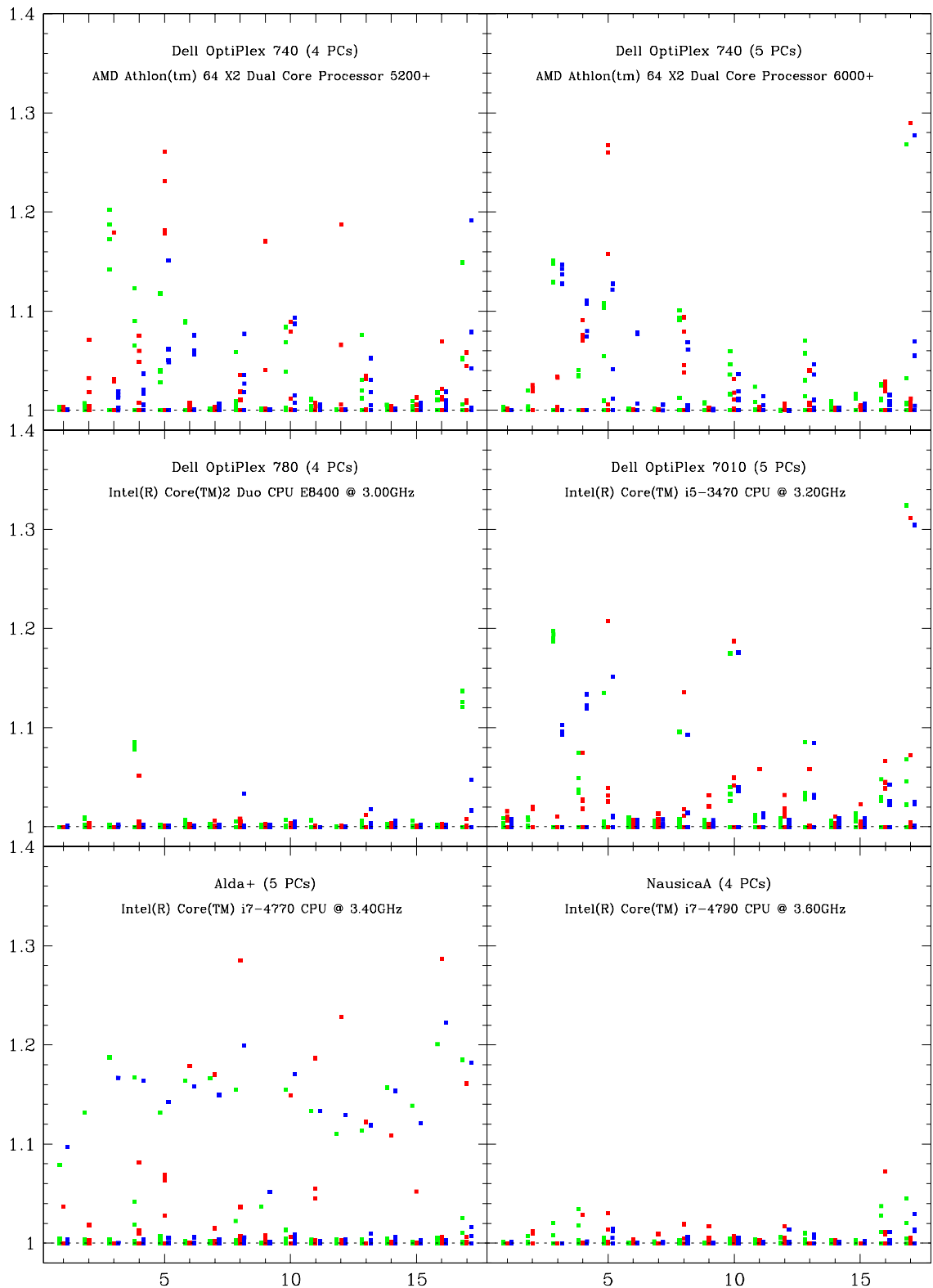}
\caption{This panel shows, for six different PC models, the scatter in
run-times after normalizing them to the best (minimum) value achieved for
each model and benchmark test. The color scheme is the same as in
Figure~\ref{Fig:comparison_same_machine}.}
\label{Fig:comparison_same_model}
\end{figure*}

The next step was to compare these results on all the PCs of a same
model. Figure~\ref{Fig:comparison_same_model} shows six examples where the
benchmark run-times are compared with the best results, that is the minimum
run-time, for all PCs of a same model. With a few exceptions, the run-times
agree to better than 20 \%. Again, we took the minimum values as
representative for that PC model, which should be a good approximation
to the theoretical limit that can be achieved on that kind of hardware.

\subsection{Comparison with reference benchmarks}

At this stage, for each PC model we have the best (that is, shortest across 
all PCs of that model) benchmark times for each test and for each compiler.

As already mentioned, for most of the powerful machines and for laptops we
have only one instance of that specific model available, so no comparisons
with other machines of the same type were possible. The benchmark data for
these machines carry then a larger uncertainty.

To provide a homogeneous set of comparisons, we took as reference the
benchmark times published on \cite{polybench},
which were measured on a "machine with a Core i5 2500k 3.30GHz processor, running
at stock speed, with 16 GBytes memory, and running 64-bit Scientific Linux 6
(a near-clone of Red Hat Enterprise Linux 6)".

The set of Figures~\ref{Fig:final_benchmarks} shows, for each PC model, the 
benchmark run-times normalized to the values listed in the above website.

\subsection{Final comparison and conclusions}

Following the scheme implemented by Polyhedron, we computed for each PC model
and each compiler the geometrical mean of the 17 execution times. The
geometrical means are then compared with those published by Polyhedron, and
shown in Figure~\ref{Fig:geometrical_mean}. As test N.\;2 (\textsl{aermod})
failed for the PGI Fortran compiler in our benchmarks, we computed the
geometrical mean excluding this test, and the Polyhedron geometrical mean for
PGI was recomputed as well excluding test N.\;2.

The graph clearly shows that CPUs in recent models have become about three
times faster than 7--8 years ago. On the other hand, laptops are in general
about as fast as a desktop PC of a same age, with the fastest laptop only
slightly slower than the fastest desktop PC.
Overall there are no significant speed differences between the three
compilers we tested, except in the "Dell Optiplex 740" desktops family 
(with AMD processors) where the Intel compiler was about 20\% slower that 
the PGI and gfortran compilers.

The results of this study will be especially useful to HTCondor users, as
they permit to restrict the list of target machines to those with the
shortest execution times, which will maximize the probability that the
submitted job are completed and not evicted, for instance, by the user 
logging in on the machine.
Furthermore, the information gathered here will help plan the upgrade of
our computing nodes.
Finally, the benchmark results will allow users to quickly assess the
performances of laptops, as compared to desktop or rack PCs, and quickly
determine whether their laptops can satisfy their computing needs.

\phantomsection
\section*{Acknowledgments} 

\addcontentsline{toc}{section}{Acknowledgments} 

We thank our colleagues in the IT Department who helped us with many small problems
related to HTCondor, especially issues with firewalls, HTCondor start-up files, etc.
A big thank is due to {\'A}ngel de Vicente, who was the first to install and 
manage HTCondor in our Institute, and the main responsible for the big popularity
it is having, in terms of usage, among our researchers.
We also thank Ubay Dorta, Justo Luna and Cristina Zurita who kindly ran the
benchmark suite on their laptops.

\phantomsection
\bibliographystyle{plain}


\begin{figure*} 
\mbox{
\centerline{
\subfigure{\includegraphics[angle=0,width=\linewidth]{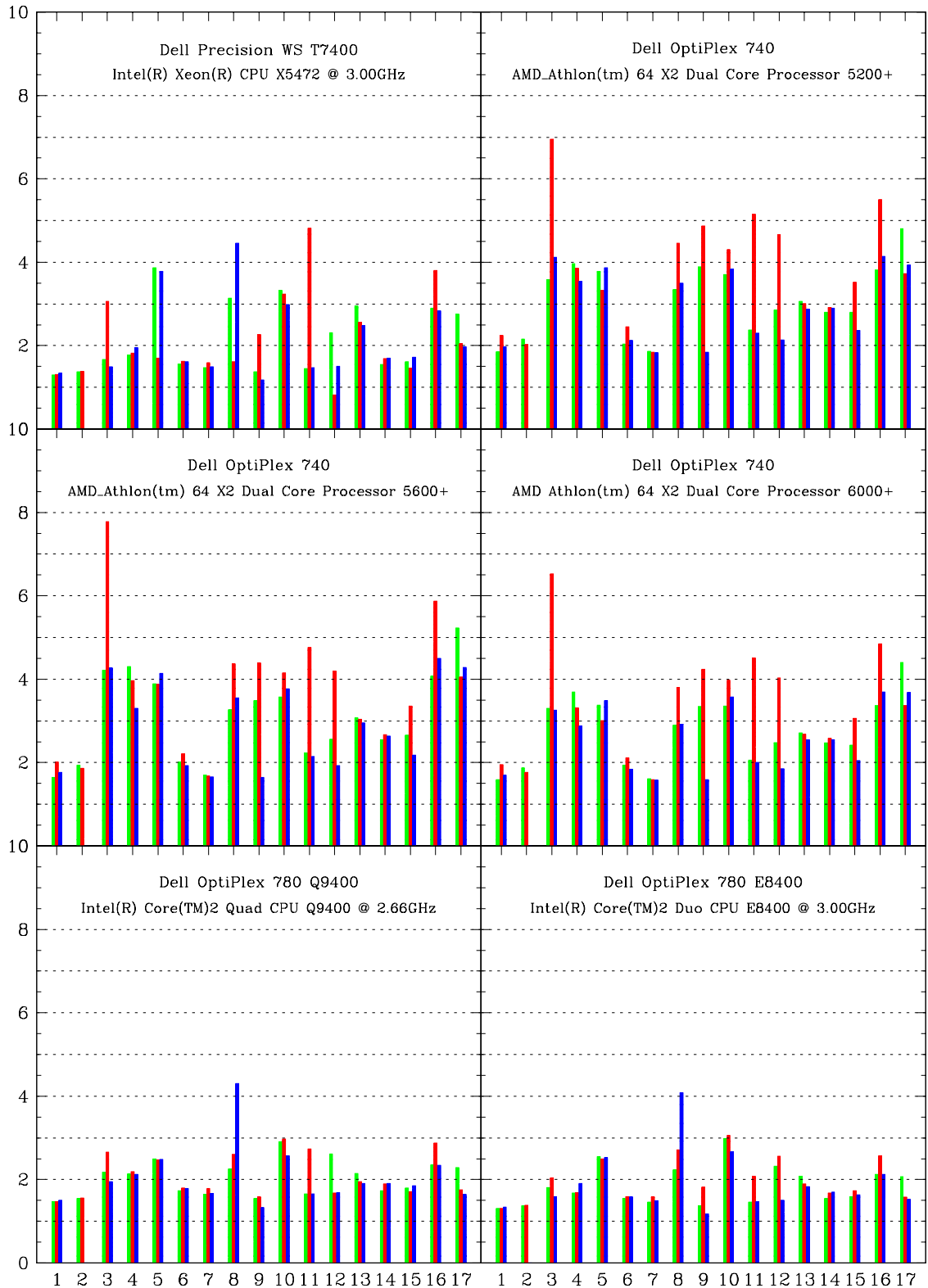}}
}}
\caption{The best (minimum) run-times obtained in our tests for desktop and
rack PCs are shown normalized to the value published by Polyhedron for a
"Sandy Bridge Intel Core i5 2500k" CPU. Again, green, red and blue indicates
the gfortran, Intel Fortran and PGI Fortran compilers respectively.}
\label{Fig:final_benchmarks}
\end{figure*}

\addtocounter{figure}{-1}

\begin{figure*} 
\mbox{
\centerline{
\subfigure{\includegraphics[angle=0,width=\linewidth]{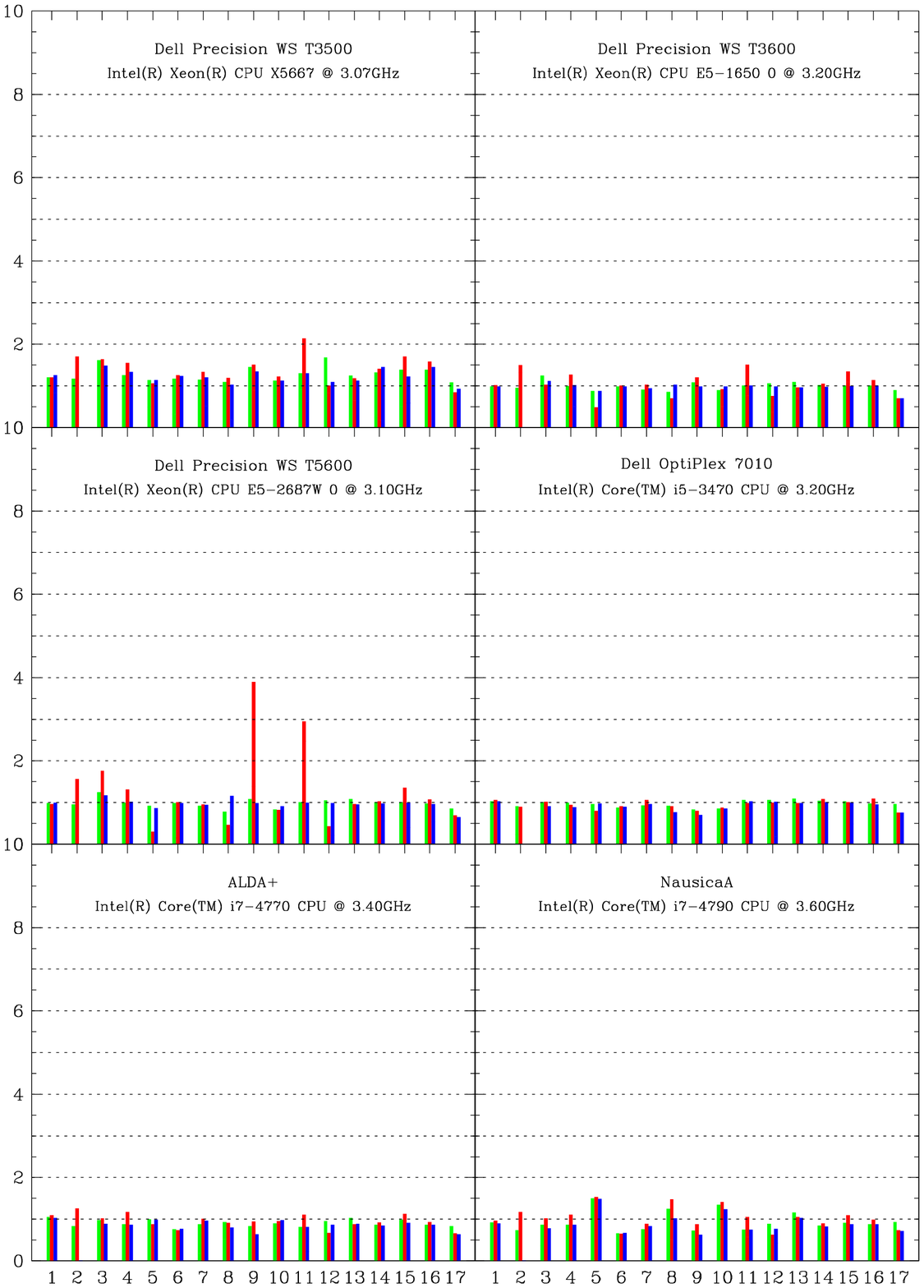}}
}}
\caption{continued}
\end{figure*}

\addtocounter{figure}{-1}

\begin{figure*} 
\mbox{
\centerline{
\subfigure{\includegraphics[angle=0,width=\linewidth]{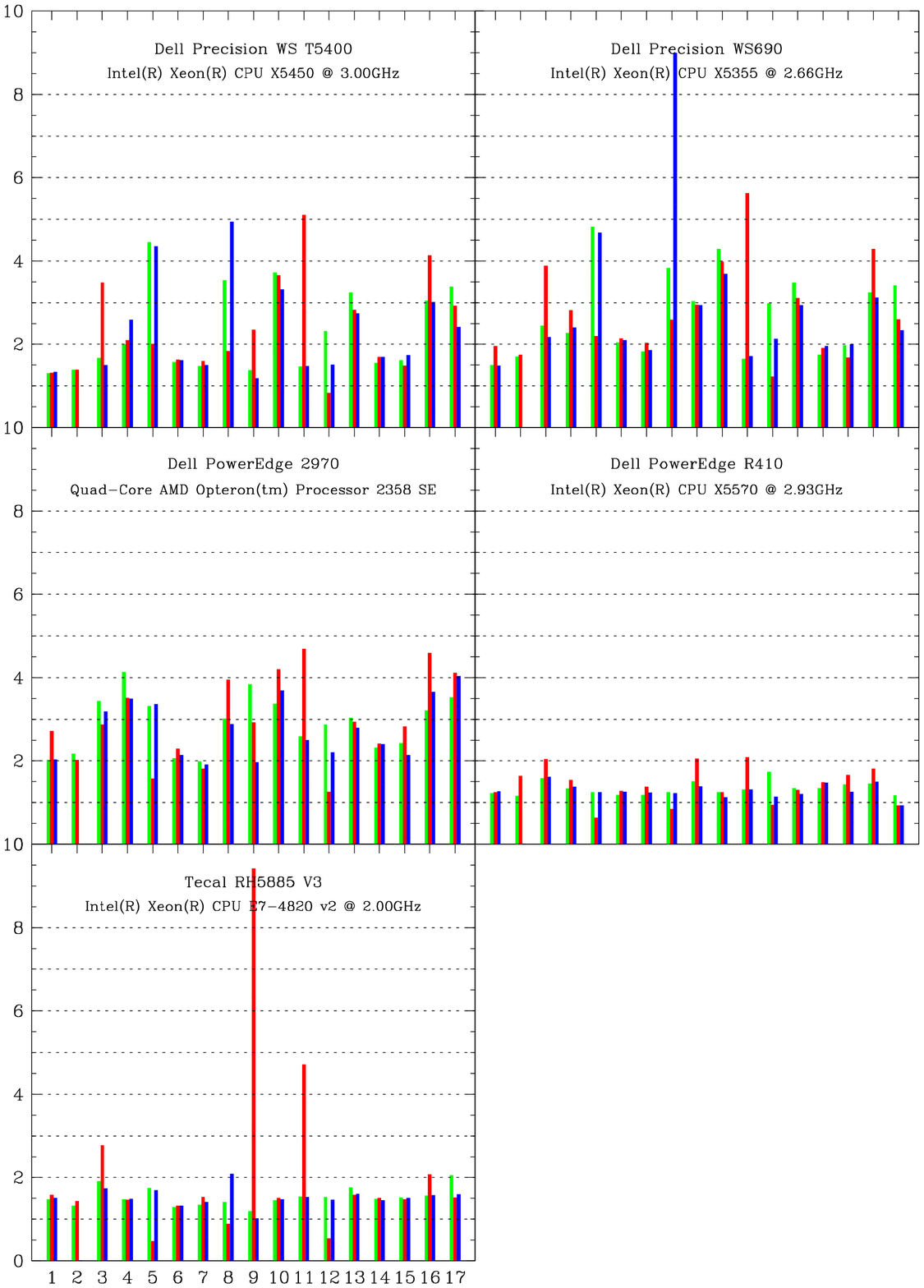}}
}}
\caption{continued}
\end{figure*}

\begin{figure*} 
\mbox{
\centerline{
\subfigure{\includegraphics[angle=0,width=\linewidth]{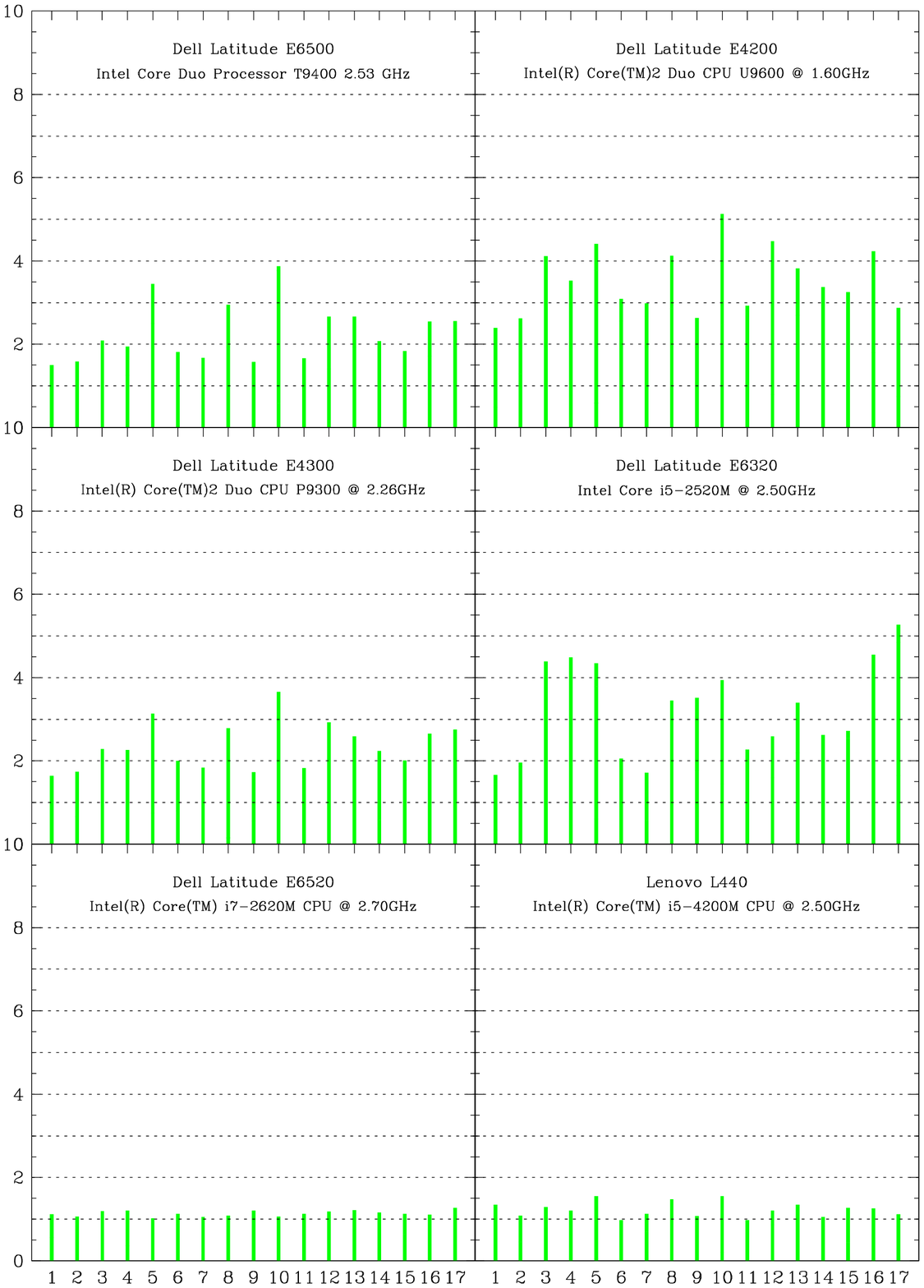}}
}}
\caption{The best (minimum) run-times obtained in our tests for laptops are
shown normalized to the value published by Polyhedron for a "Sandy Bridge
Intel Core i5 2500k" CPU. Due to license issues, only the gfortran tests were
run on laptops.}
\end{figure*}

\begin{figure*}
\mbox{
\centerline{
\subfigure{\includegraphics[angle=0,width=0.98\linewidth]{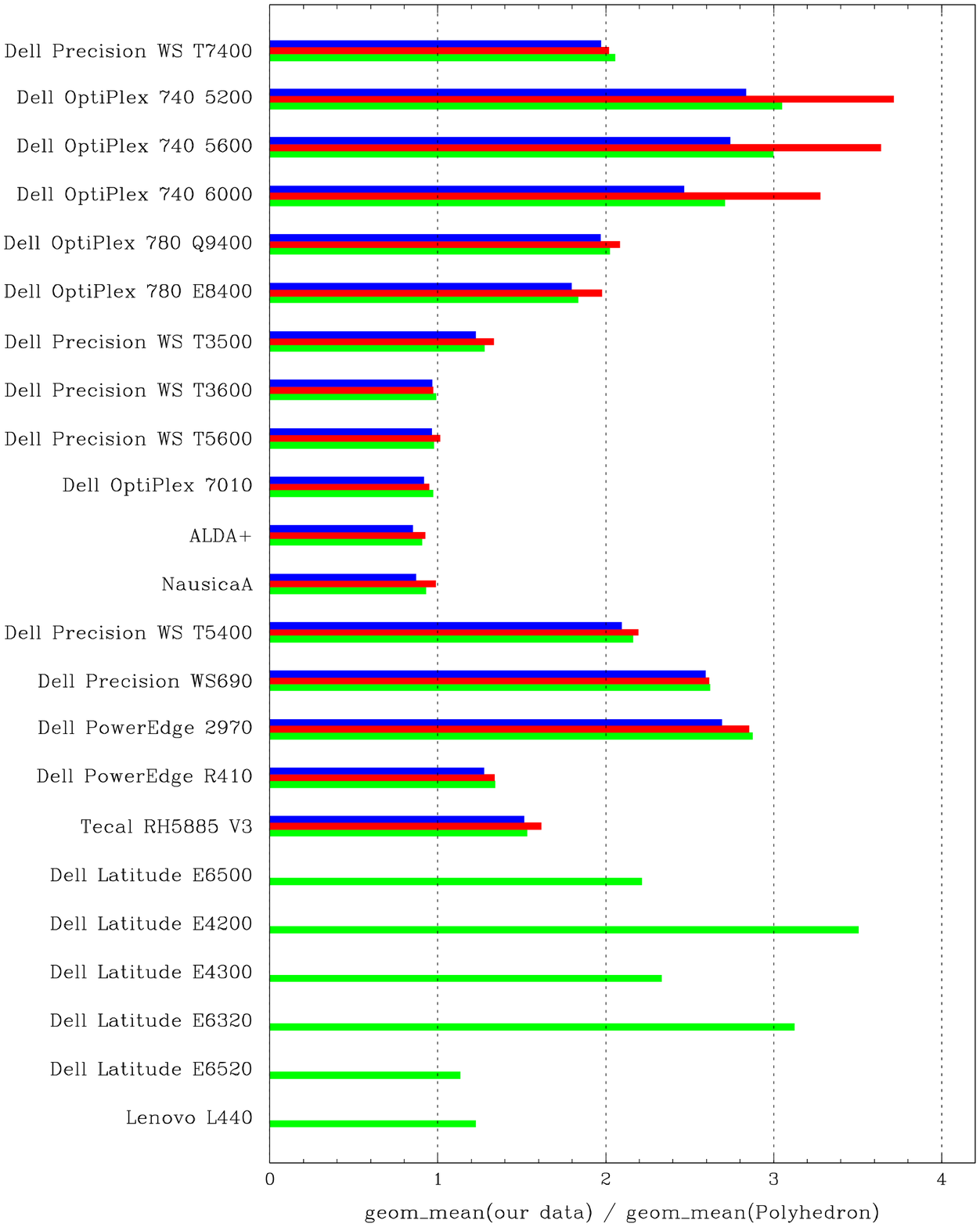}}
}}
\caption{The geometrical mean of the execution times for the 17 benchmarks are 
shown normalized to the geometrical mean values published by Polyhedron for
a Core i5 2500k 3.30GHz processor.}
\label{Fig:geometrical_mean}
\end{figure*}


\end{document}